\documentclass[12pt,letterpaper]{article}

\usepackage{amsmath}
\usepackage{amsfonts}
\usepackage{amssymb}
\usepackage{rotating}
\usepackage{color}

\newtheorem{theo}{Theorem}
\newtheorem{lem}{Lemma}

\newtheorem{defin}{Definition}
\newtheorem{prop}{Proposition}
\newtheorem{cor}{Corollary}

\newtheorem{ex}{Example}

\begin{document}

\title{\textsc{Unbeatable Imitation\thanks{We thank Carlos Al\'{o}s-Ferrer, Chen Bo, Drew Fudenberg, Alexander Matros, and John Stachurski for interesting discussions. Seminar audiences at Australian National University, Melbourne University, Monash University, UC Davis, UC San Diego, the University of Heidelberg, the University of Queensland, and at the International Conference on Game Theory in Stony Brook, 2009, contributed helpful comments.}}}
\author{Peter Duersch\thanks{Department of Economics, University of Heidelberg} \and J\"{o}rg Oechssler
\thanks{Department of Economics, University of Heidelberg, Email: oechssler@uni-hd.de} \and Burkhard C. Schipper\thanks{Department of Economics, University of California, Davis, Email: bcschipper@ucdavis.edu}}

\date{February 18, 2010}

\maketitle

\begin{abstract} We show that for many classes of symmetric two-player games, the simple decision rule ``imitate-the-best'' can hardly be beaten by any other decision rule. We provide necessary and sufficient conditions for imitation to be unbeatable and show that it can only be beaten by much in games that are of the rock-scissors-paper variety. Thus, in many interesting examples, like 2x2 games, Cournot duopoly, price competition, rent seeking, public goods games, common pool resource games, minimum effort coordination games, arms race, search, bargaining, etc., imitation cannot be beaten by much even by a very clever opponent.\\


\noindent \textbf{JEL-Classifications: } C72, C73, D43.
\end{abstract}

\thispagestyle{empty}

\newpage \setcounter{page}{1}\setlength{\baselineskip}{1.5em}

\begin{quote}
\textit{``Whoever wants to set a good example must add a grain of
foolishness to his virtue: then others can imitate and yet at the same time
surpass the one they imitate - which human beings love to do.'' Friedrich
Nietzsche}
\end{quote}

\section{Introduction}

Psychologists and behavioral economists stress the role of simple heuristics or rules for human decision making under limited computational capabilities (see Gigerenzer and Selten, 2002). While such heuristics lead to successful decisions in some particular tasks, they may be suboptimal in others. It is plausible that decision makers may cease to adopt heuristics that do worse than others in relevant situations. If various heuristics are pitted against each other in a contest, then in the long run the heuristic with the highest payoff should survive.

One of the heuristics in the contest may be a rational, omniscient, and forward looking decision rule. Even if such a rational rule is not currently among the contestants, there can always be a ``mutation'', i.e., an invention of a new rule, that enters the pool of rules. A heuristic that does very badly against such a rational rule will not be around for long as it will not belong to the top performers. Being subject to exploitation by the rational opponent in strategic situations would be an evolutionary liability. Consequently, we like to raise the following question: Is there a simple adaptive heuristic that can not be beaten even by a rational, omniscient and forward looking maximizer in large classes of economically relevant situations?

The idea for this paper emerged from a prior observation in experimental data.
In Duersch, Kolb, Oechssler, and Schipper (2010), subjects played against
computers that were programmed according to various learning algorithms in
an Cournot duopoly. On average, human subjects easily won against all of
their computer opponents with one exception: the computer following the rule
``imitate-the-best'', the rule that simply prescribes to mimic the action of the most successful player in the previous round. This suggested to us that imitation may be hard to beat even by forward--looking players.

In this paper, we prove that this holds more generally. The decision heuristic ``imitate-the-best'' is very hard to beat by \emph{any} other decision rule in large classes of symmetric two-player games that are highly relevant for economics and include games such as all symmetric 2x2 games, Cournot duopoly, Bertrand duopoly, rent seeking, public goods games, common pool resource games, minimum effort
coordination games, synergistic relationship, arms races, Diamond's search, Nash demand bargaining, etc.

We shall consider two notions of being ``unbeatable''. We call imitation ``essentially unbeatable'' if during the infinitely repeated
game against the some opponent, the opponent cannot obtain, in total, over an
infinite number of periods, a payoff difference that is more than the maximal
payoff difference for the one--period game. As a weaker notion we consider
the concept of being ``not subject to a money pump''. We say imitation is not subject to a money pump if there is a bound on the sum of payoff differences an opponent can achieve in the infinitely repeated game.

It would be intractable to explicitly consider how all possible opponents --
who may have any arbitrary decision rule -- would play against an
imitator. That is why we consider the \emph{toughest possible opponent} against the imitator in order to obtain an upper bound for how much an imitator can be beaten by any possible opponent. This toughest possible opponent clearly is a dynamic relative payoff maximizer who maximizes the sum of all future differences between her payoff
and the imitator's payoff. The relative payoff maximizer is assumed to be
infinitely patient and forward looking and never to make a mistake. More
importantly, the dynamic payoff maximizer is assumed to know that she is
matched against an imitator. That is, she knows exactly what her opponent,
the imitator, will do at all times, including the imitator's starting value. Finally, the dynamic payoff maximizer is enabled to commit to any strategy including any closed-loop strategy. Although these assumptions are certainly extreme, they make sure that if imitation cannot be beaten by this maximizer, then it cannot be beaten by any decision rule including dynamic absolute payoff maximization or any
decision rule that is more myopic or less omniscient.

Our results are as follows. We present \emph{necessary and sufficient conditions}
for imitation being subject to a money pump. The paradigmatic example for a
money pump is playing repeatedly the game rock--paper-scissors, in which, obviously, an imitator can be exploited without bounds by the maximizer. The main result
of this paper is that imitation is subject to a money pump \emph{if and only
if} the \emph{relative payoff} game in question contains a generalized
rock--paper-scissors game as a submatrix.

Since the existence of a rock--paper-scissors submatrix may be cumbersome to
check in some instances, we also provide a number of sufficient conditions
for imitation not to be subject to a money pump that are based on more
familiar concepts like quasiconcavity, generalized ordinal potentials, or quasisubmodularity/quasisupermodularity and aggregation of actions.

We also provide a number of sufficient conditions for imitation to be
essentially unbeatable like exact potentials, increasing/decreasing differences, or additive separability. One such condition is that the game is a symmetric
2x2 game. To gain some intuition for this, consider the game of
``chicken'' presented in the following
payoff matrix.
\begin{equation*}
\begin{array}{cc}
&
\begin{array}{cc}
\text{swerve} & \text{straight}
\end{array}
\\
\begin{array}{c}
\text{swerve} \\
\text{straight}
\end{array}
& \left(
\begin{array}{cc}
3,3 & 1,4 \\
4,1 & 0,0
\end{array}
\right)
\end{array}
\end{equation*}
Suppose that initially the imitator starts out with playing
``swerve''. What should a forward looking opponent do? If she decides to play ``straight'', she will earn more than the imitator today but will be copied by the imitator tomorrow. From then on, the imitator will stay with ``straight'' forever. If she decides to play ``swerve'' today, then she will earn the same as the imitator and the imitator will stay with ``swerve'' as long as the opponent stays
with ``swerve''. Suppose the opponent is a dynamic relative payoff maximizer. In that case, the dynamic relative payoff maximizer can beat the imitator at most by the maximal one-period payoff differential of 3. Now suppose the opponent maximizes the sum of her \emph{absolute} payoffs. The best an absolute payoff maximizer can do is to play swerve forever.\footnote{Payoffs are evaluated according to the over--taking criterion (see below).} In this case the imitator cannot be beaten at all as he receives the same payoff as his opponent. In either case, imitation comes very close to the top--performing heuristics and there is no need to abandon such an heuristic.

Imitate-the-best has been previously studied theoretically and experimentally mostly in Cournot oligopolies. Vega-Redondo (1997) shows that
in symmetric Cournot oligopoly with imitators, the long run outcome
converges to the competitive output if small mistakes are allowed. Huck,
Normann, and Oechssler (1999), Offerman, Potters, and Sonnemans (2002), and
Apesteguia et al. (2007, 2010) provide some experimental evidence.
Vega-Redondo's result has been generalized to aggregative quasisubmodular
games by Schipper (2003) and Al\'{o}s-Ferrer and Ania (2005). For Cournot
oligopoly with imitators and myopic best reply players, Schipper (2009)
shows that the imitators' long run average payoffs are strictly higher than
the best reply players' average payoffs.

The article is organized as follows. In the next section, we present the
model and provide formal definitions for being unbeatable. Our main result,
which provides a necessary and sufficient condition for a money pump, is
contained in Section~\ref{iff}. Sufficient conditions for imitation to be
essentially unbeatable are given in Section~\ref{essentially}. Section \ref{money pump} provides sufficient conditions for imitation not being subject
to a money pump. We finish with Section~\ref{discussion}, where we summarize
and discuss the results.

\section{Model}

We consider a symmetric\footnote{See for instance Weibull (1995, Definition 1.10).} two--player game $(X,\pi )$, in which both players are endowed with the same (finite or infinite) set of pure actions $X$ and the same bounded payoff function $\pi :X\times X\longrightarrow \mathbb{R}$, where $\pi (x, y)$ denotes the payoff to the player choosing the first argument. We will frequently make use of the following definition.
\begin{defin}[Relative payoff game]
\label{zerorel} Given a symmetric two-player game $(X, \pi)$, the relative
payoff game is $(X, \Delta)$, where the relative payoff function $\Delta: X
\times X \longrightarrow \mathbb{R}$ is defined by
\begin{equation*}
\Delta (x, y) = \pi(x, y) - \pi(y, x).
\end{equation*}
\end{defin}
Note that, by construction, every relative payoff game is a symmetric
zero-sum game since $\Delta (x,y)=-\Delta (y,x)$.

We introduce two types of players. The \emph{imitator} follows the simple rule ``imitate-the-best''. To be precise, the imitator adopts the opponent's action if and only if in the previous round the opponent's payoff was strictly higher than that of the imitator. Formally, the action of the imitator $y_{t}$ in period $t$ given the action of the other player from the previous period $x_{t-1}$ is
\begin{equation}\label{y}
y_{t}=\left\{
\begin{array}{ll}
x_{t-1}\text{ } & \text{if }\Delta (x_{t-1},y_{t-1})>0 \\
y_{t-1} & \text{else}
\end{array}
\right.
\end{equation}
for some initial action $y_{0}\in X$.

The second type we consider is a \emph{dynamic relative payoff maximizer}. The dynamic relative payoff maximizer, from now on call her the \emph{maximizer}, maximizes the sum of all future payoff differentials between her and the imitator,
\begin{equation}
D(T):=\sum_{t=0}^{T}\Delta (x_{t},y_{t}),
\end{equation}%
where $y_{t}$ is known to be given by (\ref{y}).

Since this sum may become infinite for $T\rightarrow \infty $, we assume
that the maximizer evaluates her strategies according to the \emph{overtaking-criterion} (see e.g. Osborne and Rubinstein, 1994, p. 139).
Accordingly, a sequence of relative payoffs $\{\Delta
(x_{t},y_{t})\}_{t=0}^{\infty }$ is strictly preferred to a sequence $\{\Delta (x_{t}^{\prime },y_{t}^{\prime })\}_{t=0}^{\infty }$ if $\lim_{T}\inf \sum_{t=0}^{T}(\Delta (x_{t},y_{t})-\Delta (x_{t}^{\prime},y_{t}^{\prime }))>0$.\footnote{General results on the existence of optimal over-taking strategies are developed in Leizarowitz (1996) and the literature cited therein. Here we can side-step the issue of existence because our proofs will
be constructive in the sense that (a) we construct strategies - no
matter whether optimal or not - that beat the imitator or (b) we show
that no strategy can beat the imitator.}

If we used time-averaging instead of the over-taking criterion, then the maximizer would be indifferent between a sequence of zero relative payoffs and a sequence with a finite number of strictly positive relative payoffs and zero thereafter. Yet, our aim is to pit the imitator against a maximizer who cares even about a finite number of payoff advantages. If we used time discounting instead the over-taking criterion, then the maximizer may prefer a sequence with a large payoff at the beginning over a sequence with an endless cycle of small but
positive relative payoffs. However, we believe that money pumps -- even if
small -- are a feature of irrationality that a rational opponent would take
advantage of.

It is important to realize just how extreme our assumptions regarding the
maximizer are. The maximizer is infinitely patient and forward looking and
never makes a mistake. More importantly, she is assumed to know exactly what
her opponent, the imitator, will do at all times, including the imitator's
starting value. Although these assumptions are certainly extreme, they make
sure that if imitation cannot be beaten by this maximizer, then it can not
beaten by any decision rule including dynamic absolute payoff maximization
or any decision rule that is more myopic or less omniscient.

\begin{defin}[No money pump]
\label{no money pump}We say that \emph{imitation is not subject to a money
pump} if there exists a bound $M\in \mathbb{R}_{+}$ such that for any
initial action of the imitator $y_{0}\in X$,
\begin{equation}
\lim\limits_{T\rightarrow \infty }\sup \sum_{t=0}^{T}\Delta
(x_{t},y_{t})\leq M,  \label{limsup}
\end{equation}
where $y_{t}$ is given by (\ref{y}).
\end{defin}

That is, imitation is not subject to a money pump if it can be beaten only
by a finite amount although the game between the imitator and the maximizer
runs for an infinite number of periods. In some cases we can show that
imitation can in fact not be beaten by more than the payoff differential
from a single period.

\begin{defin}[Essentially unbeatable] We say that \emph{imitation is essentially unbeatable} if it can be beaten in total by at most the maximal one-period payoff differential, i.e., if $M$ in inequality~(\ref{limsup}) is at most $\hat{\Delta}:=\max_{x,y}\Delta
(x,y) $.
\end{defin}

As in previous studies of imitation (see e.g. Al\'{o}s-Ferrer and Ania, 2005; Schipper, 2003; Vega-Redondo, 1997), the concept of a finite population evolutionary stable strategy (Schaffer, 1988, 1989) plays a prominent role in our analysis.

\begin{defin}[fESS]
\label{fESS} An action $x^* \in X$ is a \emph{finite population evolutionary
stable strategy (fESS)} of the game $(X, \pi)$ if
\begin{equation}\label{fESSformula}
\pi(x^*, x) \geq \pi(x, x^*) \mbox{ for all } x \in X.
\end{equation}
\end{defin}
In terms of the relative payoff game, inequality~(\ref{fESSformula}) is
equivalent to
\begin{equation*}
\Delta (x^*, x) \geq 0 \mbox{ for all } x \in X.
\end{equation*}

Already Schaffer (1988, 1989) observed that the fESS of the game $(X, \pi)$
and the symmetric pure Nash equilibria of the relative payoff game $(X,\Delta )$ coincide.

\section{A Necessary and Sufficient Condition for a Money Pump\label{iff}}

The game rock--paper-scissors is the paradigmatic example for how an
imitator can be exploited without bounds by the maximizer. In our
terminology, imitation is subject to a money pump.

\begin{ex}[Rock-Paper-Scissors]
\label{RPS} Consider the well known rock-paper-scissors game.\footnote{In the following, we will represent symmetric payoff matrices by the matrix of the row player's payoffs only.}
\begin{equation*}
\begin{array}{cc}
&
\begin{array}{ccc}
R~ & ~P & ~S%
\end{array}
\\
\begin{array}{c}
R \\
P \\
S%
\end{array}
& \left(
\begin{array}{ccc}
0 & -1 & 1 \\
1 & 0 & -1 \\
-1 & 1 & 0%
\end{array}%
\right)%
\end{array}%
\end{equation*}
Clearly, if the imitator starts for instance with R, then the dynamically
optimal strategy of the maximizer is the cycle P-S-R... In this way, the
maximizer wins in every period and the imitator loses in every period. Over
time, the payoff difference will grow without bound in favor of the
maximizer.
\end{ex}

We can generalize Example~\ref{RPS} by noting that the crucial feature of
the example is that the maximizer can find for each action of the imitator
an action which yields her a strictly positive relative payoff.

\begin{defin}[Generalized Rock-Paper-Scissors Matrix]\label{grps_matrix copy(1)} A symmetric zero-sum game $(X, \pi)$ is called a generalized rock-paper-scissors matrix if for each column there exists a row with a strictly positive payoff to player 1.
\end{defin}

It should be fairly obvious that if a zero--sum game contains somewhere a
submatrix that is a generalized rock-paper-scissors matrix, then this is
sufficient for a money pump as the maximizer can make sure that the process
cycles forever in this submatrix. What is probably less obvious is that the
existence of such a submatrix is also necessary for a money pump.

\begin{defin}[Generalized Rock-Paper-Scissors Game]
A symmetric zero-sum game $(X, \pi)$ is called a generalized
rock-paper-scissors game if it contains a submatrix $(\bar{X}, \bar{\pi})$
with $\bar{X} \subseteq X$ and $\bar{\pi}(x, y) = \pi (x, y)$ for all $x, y \in
\bar{X}$, and $(\bar{X}, \bar{\pi})$ is a generalized rock-paper-scissors
matrix.
\end{defin}

This leads us to our main result.

\begin{theo}\label{characterization} Imitation is subject to a money pump in the finite symmetric game $(X, \pi)$ if and only if its relative payoff game $(X, \Delta)$ is a generalized rock-paper-scissors game.
\end{theo}

Recall that according to Definition~\ref{no money pump}, imitation is
subject to a money pump if there exists some initial condition $y_{0} \in X$
such that inequality~(\ref{limsup}) is violated. The proof of the theorem follows
directly from the following three lemmata. We use the following preliminary
observation repeatedly in the analysis.

\begin{lem}\label{neverworse} Consider a symmetric game $(X, \pi)$ with its relative payoff game $(X, \Delta)$. The maximizer will never choose an action $x_{t}$ such that $\Delta(x_{t},y_{t}) < 0$.
\end{lem}

\noindent \textsc{Proof. } Suppose to the contrary that the maximizer
chooses an action $x_{t}$ such that $\Delta (x_{t},y_{t})<0$. Then in period
$t+1$, the imitator will not imitate her period $t$ action. But then, she
could improve her relative payoff in $t$ by setting the same action in $t$
as the imitator without influencing the actions of the imitator in period $t+1$ or any other future period.\hfill $\Box$\newline

Given a symmetric two-player game $(X,\pi )$ and its relative
payoff game $(X, \Delta)$, a path in the action space $X \times X$ is a
sequence of action profiles $(x_{0},y_{0}),(x_{1},y_{1}),...$. A path is
constant if $(x_{t},y_{t}) = (x_{t+1},y_{t+1})$ for all $t=0, 1, ...$.
Otherwise, the path is called non--constant. A non--constant finite path $(x_{0}, y_{0}), ..., (x_{n}, y_{n})$ is a \emph{cycle} if $(x_{0}, y_{0}) = (x_{n}, y_{n})$. A cycle is an \emph{imitation cycle} if for all $(x_t, y_t)$ and $(x_{t+1}, y_{t+1})$ on the path of the cycle $\Delta(x_{t},y_{t}) > 0$ and $y_{t+1}=x_{t}$. Along an imitation cycle, one player always obtains a strictly
positive relative payoff and the other player mimics the action of the first
player in the previous round. Thus, an imitation cycle never contains an
action profile on the diagonal of the payoff matrix.

\begin{lem}\label{imitation_cycle} For any finite symmetric game $(X, \pi)$,  imitation is subject to a money pump if and only if there exists an
imitation cycle.
\end{lem}

\noindent \textsc{Proof. } Consider a finite symmetric game $(X, \pi)$ and its relative payoff game $(X, \Delta)$. We show first that if imitation is subject to a money pump, then there is a imitation cycle. Since the game is finite, there can not be infinitely many strictly positive relative payoff improvements unless there
is a cycle. To show that such a cycle implies an imitation cycle, suppose by contradiction that there exists a period $t$ such that $\Delta(x_{t}, y_{t}) \leq 0$. W.l.o.g. assume that $\Delta(x_{t+1}, y_{t+1}) > 0$. This is w.l.o.g. because we assumed a money pump. By equation~(\ref{y}) the imitator will not imitate in $t+1$ the previous period's action of the maximizer, i.e., $y_{t+1} = y_t$. But then the maximizer could strictly improve the sum of her relative payoffs already in $t$ by setting $x_{t} = x_{t+1}$. Thus, there must be a cycle with $\Delta (x_{t}, y_{t}) > 0$ for all $t$. The decision rule of the imitator then requires that $y_{t+1}=x_{t}$ for all $t$, which proves that such a cycle is an imitation cycle.

The converse is trivial. \hfill $\Box$

\begin{lem} Consider a finite symmetric game $(X, \pi)$ with its relative payoff game $(X, \Delta)$. $(X,\Delta )$ is a generalized rock-paper-scissors game if and only if there exists an imitation cycle.
\end{lem}

\noindent \textsc{Proof. } \textquotedblleft $\Leftarrow $\textquotedblright
: If there exists an imitation cycle in $(X,\Delta )$, let $\bar{X}$ be the
orbit of the cycle, i.e., all actions of $X$ that are played along the
imitation cycle. For each action (i.e., column) $y \in \bar{X}$, there exists
an action (i.e., row) $x\in \bar{X}$ such that $\Delta (x,y) > 0$. Hence, $(\bar{X},\bar{\Delta})$, where $\bar{\Delta}$ is defined by $\bar{\Delta}(x, y) = \Delta (x,y)$ for all $x, y \in \bar{X}$, is a generalized rock-paper-scissors submatrix. Thus, $(X, \Delta)$ is a generalized rock-paper-scissors game.

\textquotedblleft $\Rightarrow $\textquotedblright : If the relative payoff
game $(X,\Delta )$ is a generalized rock-paper-scissors game, then it
contains a generalized rock-paper-scissors submatrix $(\bar{X},\bar{\Delta})$%
. That is, for each column of the matrix game $(\bar{X},\bar{\Delta})$ there
exists a row with a strictly positive relative payoff to player 1. Let the
initial action of the imitator $y$ be contained in $\bar{X}$. If the
maximizer selects such a row $x \in \bar{X}$ for which she earns a strict
positive relative payoff, i.e., $\Delta(x, y) > 0$, then she will be imitated by the imitator in the next period. Yet, at the next round, when the imitator plays $x$, the maximizer has another action $x^{\prime} \in \bar{X}$ with a strictly
positive relative payoff, i.e., $\Delta(x', x) > 0$. Thus the imitator will imitate her in the following period. More generally, for each action $y\in \bar{X}$ of the
imitator, there is another action $x\in \bar{X}$, $x\neq y$ of the maximizer
that earns the latter a strictly positive relative payoff. Since $\bar{X}$ is
finite, such a sequence of actions must contain a cycle. Moreover, we just
argued that $\Delta (x_{t},y_{t})>0$ and $y_{t+1}=x_{t}$ for all $t$. Thus,
it is an imitation cycle. \hfill $\Box$\\

Theorem~\ref{characterization} is used to obtain an interesting necessary
condition for imitation being not subject to a money pump.

\begin{prop} Let $(X, \pi)$ be a finite symmetric game with its relative payoff game $(X, \Delta)$. If $(X, \Delta)$ has no pure saddle point, then
imitation is subject to a money pump.
\end{prop}

\noindent \textsc{Proof. } By Theorem 1 in Duersch, Oechssler, and Schipper
(2010), $(X, \Delta)$ has no symmetric pure saddle point if and
only if it is a generalized rock-paper-scissors matrix. Thus, if $(X, \Delta)$ has no symmetric pure saddle point, then it is a generalized rock-paper-scissors game. Hence, by Theorem~\ref{characterization} imitation
is subject to a money pump. \hfill $\Box$

\begin{cor} If the finite symmetric game $(X, \pi)$ has no fESS, then imitation is subject to a money pump.
\end{cor}

In other words, the existence of a fESS is a necessary condition for
imitation not being subject to a money pump. The reason for the existence of
a fESS not being sufficient is that there could be a generalized rock-paper-scissors submatrix of the game (``disjoint'' from the fESS profile) that gives rise to an imitation cycle. If the initial action of the imitator lies within the action set corresponding to this submatrix, then imitation is subject to a money pump.

Since the relative payoff game of a symmetric zero-sum game is a generalized
rock-paper-scissors game if and only if the underlying symmetric zero-sum game is a
generalized rock-paper-scissors game, we obtain from Theorem~\ref{characterization} the following corollary.

\begin{cor} Imitation is subject to a money pump in the finite symmetric zero-sum game $(X, \pi)$ if and only if $(X, \pi)$ is a generalized rock-paper-scissors game.
\end{cor}

\section{Sufficient Conditions for Essentially Unbeatable\label{essentially}}

\subsection{Symmetric 2x2 games\label{section 2x2}}

In this section, we extend the ``chicken'' example of the introduction to all symmetric 2x2 games. Note that the relative payoff game of any symmetric 2x2 game cannot be a generalized rock--paper--scissors matrix since latter must by a symmetric zero--sum game. If one of the row player's off-diagonal relative payoffs is $a > 0$, then the other must be $-a$ violating the definition of generalized rock-paper-scissors matrix. Thus Theorem~\ref{characterization} implies that for any symmetric 2x2 game imitation is not subject to a money pump. We can strengthen the result to imitation being essentially unbeatable.

\begin{prop}\label{2x2} In any symmetric 2x2 game, imitation is essentially unbeatable.
\end{prop}

\noindent \textsc{Proof. } Let $X = \{x, x^{\prime }\}$. Consider a period $t$ in which the maximizer achieves a strictly positive relative payoff, $\Delta (x, x^{\prime }) > 0$. (If no such period $t$ in which the maximizer
achieves a strictly positive relative payoff exists, then trivially
imitation is essentially unbeatable.) By definition, $\Delta (x, x^{\prime
}) \leq \hat{\Delta}$. Since $\Delta (x, x^{\prime }) > 0$, the imitator
imitates $x$ in period $t+1$. For there to be another period in which the
maximizer achieves a strictly positive relative payoff, it must hold that $\Delta (x^{\prime }, x) > 0$, which because of symmetry yields a contradiction as $\Delta (x^{\prime }, x) = -\Delta(x, x^{\prime })$. Thus there can be at most one period in which the maximizer achieves a strictly positive relative payoff.\hfill $\Box$ \newline

Note that ``Matching pennies'' is not a counter-example since it is not symmetric.

\subsection{Additively Separable Relative Payoff Functions}

Next, we consider relative payoff functions that are additively separable in
the players' actions. In this class of symmetric games, imitation is also
essentially unbeatable. While additive separability may appear to be
restrictive, we will show below that there is a fairly large number of
important examples that fall into this class.

\begin{defin}[Additive Separable] We say that a relative payoff function $\Delta $ is additively separable if $\Delta (x,y)=f(x)+g(y)$ for some functions $f,g:X\longrightarrow \mathbb{R}$.
\end{defin}

\begin{prop}\label{separable} Let $(X, \pi)$ be a symmetric game with its relative payoff game $(X, \Delta)$. If $X$ is compact and the relative payoff function $\Delta$ is upper semicontinuous and additively separable, then imitation is essentially unbeatable.
\end{prop}

\noindent \textsc{Proof. } Since $\Delta $ is separable, we have that for
all $x^{\prime \prime },x^{\prime },x\in X,$
\begin{equation*}
\Delta (x^{\prime \prime },x)-\Delta (x^{\prime },x)=\Delta (x^{\prime
\prime },x^{\prime })-\Delta (x^{\prime },x^{\prime }),
\end{equation*}
which is equivalent to
\begin{equation}\label{small_or_large steps}
\Delta (x^{\prime \prime },x)=\Delta (x^{\prime \prime },x^{\prime })+\Delta
(x^{\prime },x)
\end{equation}
because $\Delta (x^{\prime },x^{\prime })=0$ since the relative payoff game
is a symmetric zero--sum game.

By induction, equation~(\ref{small_or_large steps}) implies that one large
step is just as profitable as \emph{any} number of steps. Suppose three
steps were optimal for the maximizer. By equation~(\ref{small_or_large steps}) the maximizer is no worse off by merging two of the three steps to one
larger step. Applying equation~(\ref{small_or_large steps}) again yields the
claim.

Thus, if a fESS exist, then the maximizer cannot do better than jumping directly to a fESS $x^{\ast}$ since for all $x, y \in X$,
\begin{equation*}
\Delta (x^{\ast },y)=\Delta (x^{\ast },x)+\Delta (x,y)\geq \Delta (x,y),
\end{equation*}
where the equality follows from equation~(\ref{small_or_large steps}) and
the inequality from the definition of fESS. If the inequality is strict,
then once the maximizer has chosen $x^{\ast}$, the imitator will follow and
remain there for ever. Otherwise, if the inequality holds with equality then
the maximizer can not improve further his relative payoff.

Finally, in Duersch, Oechssler, and Schipper (2010, Corollary 6), we show that if $X$ is compact and $\Delta$ upper semicontinuous and additively separable, then a
fESS of $(X,\pi )$ indeed exists.\hfill $\Box$\newline

The following corollary follows directly from Proposition~\ref{separable}
and may be useful in applications.

\begin{cor}\label{separable2} Consider a game $(X, \pi)$ with a compact action set $X$ and a payoff function that can be written as $\pi(x,y)=f(x) + g(y) + a(x,y)$
for some continuous functions $f, g : X \longrightarrow \mathbb{R}$ and a symmetric function $a: X \times X \longrightarrow \mathbb{R}$ (i.e., $a(x,y)=a(y,x)$ for all $x,y\in X$). Then imitation is essentially unbeatable.
\end{cor}

Properties such as increasing or decreasing differences are often useful for proving the existence of pure equilibria and convergence of learning processes.

\begin{defin} Let $X$ be a totally ordered set. A (relative) payoff function $\Delta$ has decreasing (resp. increasing) differences on $X\times X$ if for all $x^{\prime \prime },x^{\prime },y^{\prime \prime },y^{\prime }\in X$ with $x^{\prime \prime }>x^{\prime }$ and $y^{\prime \prime }>y^{\prime }$,
\begin{equation}\label{decreasing diff}
\Delta (x^{\prime \prime },y^{\prime \prime })-\Delta (x^{\prime },y^{\prime
\prime })\leq (\geq )\Delta (x^{\prime \prime },y^{\prime })-\Delta
(x^{\prime },y^{\prime }).
\end{equation}
$\Delta $ is a valuation if it has both decreasing and increasing
differences.
\end{defin}

Our original intent was to study the consequences of $\Delta (x,y)$ having
either increasing or decreasing differences. However, in Duersch, Oechssler,
and Schipper (2010, Lemma 1) we show that for all symmetric two-player zero-sum
games, increasing differences is equivalent to decreasing differences. By
Topkis (1998, Theorem 2.6.4.), a function is additively separable on a totally ordered set $X$ if and only if it is a valuation. Hence, we have the following corollary to Proposition~\ref{separable}.

\begin{cor}\label{order} Let $(X, \pi)$ be a finite symmetric game with its relative payoff game $(X, \Delta)$. If $X$ is a totally ordered set and $\Delta $ has increasing or decreasing differences or is a valuation, then imitation is essentially unbeatable.
\end{cor}

Br\^{a}nzei, Mallozzi, and Tijs (2003, Theorem 1) show that a zero-sum game is an exact potential game if and only if it is additively separable.
Thus, Proposition~\ref{separable} implies that imitation is essentially
unbeatable in exact potential games. The following notion is due to Monderer and Shapley (1996).

\begin{defin}[Exact potential games]\label{exact_pot} The symmetric game $(X, \pi)$ is an exact potential game if there exists an exact potential function $P:X \times X\longrightarrow \mathbb{R}$ such that for all $y \in X$ and all $x,x^{\prime }\in X$,
\begin{eqnarray*}
\pi(x,y)-\pi (x^{\prime}, y) & = & P(x,y)-P(x^{\prime },y), \\
\pi(x,y)-\pi (x^{\prime}, y) & = & P(y,x)-P(y,x^{\prime}).
\end{eqnarray*}
\end{defin}

\begin{cor}\label{exact} Let $(X,\pi )$ be a finite symmetric game with its relative payoff game $(X,\Delta )$. If $(X,\Delta )$ is an exact potential game, then imitation is essentially unbeatable.
\end{cor}

All of the following examples follow from Proposition~\ref{separable} or
Corollary~\ref{separable2}. They demonstrate that the assumption of additively separable relative payoffs is not as restrictive as may be thought at first
glance.

\begin{ex}[Cournot Duopoly with Linear Demand]\label{linear_Cournot} Consider a Cournot duopoly given by the symmetric payoff function by $\pi (x,y)=x(b-x-y)-c(x)$ with $b>0$. Since $\pi(x,y)$ can be written as $\pi (x,y)=bx-bx^{2}-c(x)-xy$, Corollary~\ref{separable2} applies, and imitation is essentially unbeatable.
\end{ex}

The following example with strategic complementarities shows that the result
is not restricted to strategic substitutes.

\begin{ex}[Bertrand Duopoly with Product Differentiation]\label{Bertand} Consider a differentiated duopoly with constant marginal costs, in which firms 1 and 2 set prices $x$ and $y$, respectively. Firm 1's profit function is given by $\pi (x,y)=(x-c)(a+by-\frac{1}{2}x)$, for $a>0$, $b \in [0,1/2)$. Since $\pi (x,y)$ can be written as $\pi (x,y) = ax-ac+\frac{1}{2}cx-\frac{1}{2}x^{2}-bcy+bxy$, Corollary~\ref{separable2} applies, and imitation is essentially unbeatable.
\end{ex}

\begin{ex}[Public Goods]\label{public_goods} Consider the class of symmetric public good games defined by $\pi (x,y)=g(x,y)-c(x)$ where $g(x,y)$ is some symmetric monotone increasing benefit function and $c(x)$ is an increasing cost function.
Usually, it is assumed that $g$ is an increasing function of the sum of
provisions, that is the sum $x+y$. Various assumptions on $g$ have been
studied in the literature such as increasing or decreasing returns. In any
case, Corollary~\ref{separable2} applies, and imitation is essentially unbeatable.
\end{ex}

\begin{ex}[Common Pool Resources]\label{CPR} Consider a common pool resource game with two appropriators. Each appropriator has an endowment $e>0$ that she can invest in an outside activity with marginal payoff $c>0$ or into the common pool resource. $x \in X\subseteq \lbrack 0,e]$ denotes the maximizer's investment into the common pool resource (likewise $y$ denotes the imitator's investment). The return from investment into the common pool resource is $\frac{x}{x+y}(a(x+y)-b(x+y)^{2})$, with $a,b>0$. So the symmetric payoff function is given by $\pi(x,y)=c(e-x)+\frac{x}{x+y}(a(x+y)-b(x+y)^{2})$ if $x,y>0$ and $ce$ otherwise (see Walker, Gardner, and Ostrom, 1990). Since $\Delta(x,y)=(c(e-x)+ax-bx^{2})-(c(e-y)+ay-by^{2})$, Proposition~\ref{separable} implies that imitation is essentially unbeatable.
\end{ex}

\begin{ex}[Minimum Effort Coordination]\label{minimum_effort} Consider the class of minimum effort games given by the symmetric payoff function $\pi (x,y)=\min \{x,y\}-c(x)$ for some cost function $c$ (see Bryant, 1983 and Van Huyck, Battalio, and Beil, 1990). Corollary~\ref{separable2} implies that imitation is essentially unbeatable.
\end{ex}

\begin{ex}[Synergistic Relationship] Consider a synergistic relationship among two individuals. If both devote more effort to the relationship, then they are both better off, but for any given effort of the opponent, the return of the player's effort first increases and then decreases. The symmetric payoff function is given by $\pi(x, y) = x (c + y - x)$ with $c > 0$ and $x, y \in X \subset \mathbb{R}_+$ with $X$ compact (see Osborne, 2004, p. 39). Corollary~\ref{separable2} implies that imitation is essentially unbeatable.
\end{ex}

\begin{ex}[Arms Race]\label{arms_race} Consider two countries engaged in an arms race (see e.g. Milgrom and Roberts, 1990, p. 1272). Each player chooses a level of arms in a compact totally ordered set $X$. The symmetric payoff function is given by $\pi(x,y)=h(x-y)-c(x)$ where $h$ is a concave function of the difference between both players' level of arms, $x-y$, satisfying $h(x-y)=-h(y-x)$. By Proposition~\ref{separable} imitation is essentially unbeatable.
\end{ex}

\begin{ex}[Diamond's Search]\label{diamond_search} Consider two players who exert effort searching for a trading partner. Any trader's probability of finding another particular trader is proportional to his own effort and the effort by the other. The payoff function is given by $\pi (x,y)=\alpha xy-c(x)$ for $\alpha > 0$ and $c$ increasing (see Milgrom and Roberts, 1990, p. 1270). The relative payoff game of this two-player game is additively separable. By Proposition~\ref{separable} imitation is essentially unbeatable.
\end{ex}

A natural question is whether additive separability of relative payoffs is also necessary for imitation to be essentially unbeatable. The following counter-example shows that this is not the case.

\begin{ex}[Coordination game with outside option]\label{gopotential} Consider the following coordination game with an outside option ($C$) for both players of not participating (left matrix).
\begin{equation*}
\begin{array}{ccc}
\pi =
\begin{array}{cc}
&
\begin{array}{ccc}
A & B & C
\end{array}
\\
\begin{array}{c}
A \\
B \\
C
\end{array}
& \left(
\begin{array}{ccc}
4 & -1 & 0 \\
2 & 3 & 0 \\
0 & 0 & 0
\end{array}
\right)%
\end{array}
&  & \Delta =
\begin{array}{cc}
&
\begin{array}{ccc}
A & B & C
\end{array}
\\
\begin{array}{c}
A \\
B \\
C
\end{array}
& \left(
\begin{array}{ccc}
0 & -3 & 0 \\
3 & 0 & 0 \\
0 & 0 & 0
\end{array}
\right)
\end{array}
\end{array}
\end{equation*}
Note that the relative payoff game $\Delta$ (right matrix) does not have constant differences. E.g., $\Delta(A, B) - (B, B) = -3 \neq \Delta(A, C) - \Delta(B, C) = 0$. Thus, by Topkis (1998, Theorem 2.6.4.), it is not additively separable. Yet, imitation is essentially unbeatable. If the imitator's initial action is $A$, the maximizer can earn at most a relative payoff differential of $3$ after which the imitator adjusts and both earn zero from there on. For other initial actions of the imitator, the maximal payoff difference is at most $0$.
\end{ex}

\section{Sufficient Conditions for No Money Pump\label{money pump}}

\subsection{Relative Payoff Games with Potentials}

Potential functions are often useful for obtaining results on convergence of
learning algorithms to equilibrium, existence of pure equilibrium, and
equilibrium selection. In the previous section, we have shown that if the relative payoff game is an exact potential game, then imitation is essentially unbeatable. It is natural to explore the implications of more general notions of potentials. Besides exact potential games (see Definition~\ref{exact_pot}), the following notions of potential games were introduced by Monderer and Shapley (1996).

\begin{defin}[Potential games] The symmetric game $(X,\pi )$ is
\begin{itemize}
\item[(W)] a weighted potential game if there exists a weighted potential
function $P:X\times X\longrightarrow \mathbb{R}$ and a weight $w \in
$ $\mathbb{R}_{+}$ such that for all $y\in X$ and all $x,x^{\prime }\in X$,
\begin{eqnarray*} \pi(x, y) - \pi(x^{\prime }, y) & = & w (P(x,y) - P(x^{\prime}, y)), \\ \pi(x, y) - \pi(x^{\prime }, y) & = & w \left(P(y,x) - P(y,x^{\prime })\right).
\end{eqnarray*}

\item[(O)] an ordinal potential game if there exists an ordinal potential
function $P:X\times X\longrightarrow \mathbb{R}$ such that for all $y\in X$
and all $x,x^{\prime }\in X$,
\begin{eqnarray*} \pi (x,y)-\pi (x^{\prime },y) > 0 & \mbox{ if and only if } & P(x,y)-P(x^{\prime },y) > 0, \\ \pi (x,y)-\pi (x^{\prime },y) > 0 & \mbox{ if and only if } & P(y,x)-P(y,x^{\prime }) > 0.
\end{eqnarray*}

\item[(G)] a generalized ordinal potential game if there exists a generalized ordinal potential function $P:X\times X\longrightarrow \mathbb{R}$ such that for all $y\in X$ and all $x,x^{\prime }\in X$,
\begin{eqnarray*} \pi (x,y)-\pi (x^{\prime },y) > 0 & \mbox{ implies } & P(x,y)-P(x^{\prime },y) > 0, \\ \pi (x,y)-\pi (x^{\prime },y) > 0 & \mbox{ implies } & P(y,x)-P(y,x^{\prime }) > 0.
\end{eqnarray*}
\end{itemize}
\end{defin}

Note that every exact potential game is a weighted potential game, every
weighted potential game is an ordinal potential game, and every ordinal
potential game is a generalized ordinal potential game. Monderer and Shapley
(1996, Lemma 2.5 and the first paragraph on p. 129) show that any finite
strategic game admitting a generalized ordinal potential possesses a pure Nash equilibrium. Thus, if $(X,\pi )$ is a finite symmetric game with relative playoff game $(X,\Delta )$ and the latter is an exact, weighted, ordinal or generalized ordinal potential game, then $(X, \pi)$ possesses a fESS.

A sequential path in the action space $X \times X$ is a sequence $(x_0, y_0), (x_1, y_1), ...$ of profiles $(x_t, y_t) \in X \times X$ such that for all $t = 0, 1, ...$, the action profiles $(x_t, y_t)$ and $(x_{t+1}, y_{t+1})$ differ in exactly one player's action. A sequential path is a strict improvement path if for each $t = 0, 1, ...$, the player who switches her action at $t$ strictly improves her payoff. A finite sequential path $(x_0, y_0), ..., (x_m, y_m)$ is a \emph{strict improvement cycle} if it is a strict improvement path and $(x_0, y_0) = (x_m, y_m)$.

\begin{lem}\label{improvement_imitation} If $(X, \Delta)$ does not contain a strict
improvement cycle, then it does not contain an imitation cycle.\footnote{Ania (2008, Proposition 3) presents a similar result according to which if all players are imitators and imitation is payoff improving, then fESS implies Nash equilibrium action. This is different from Lemma~\ref{improvement_imitation} as we consider one maximizer and one imitator and focus on the relationship between relative payoff games that possess a generalized ordinal potential and imitation cycles.}
\end{lem}

\noindent \textsc{Proof. } We prove the contrapositive. I.e., if $(X,
\Delta)$ contains an imitation cycle, then it contains a strict
improvement cycle. Let $(x_0, y_0), ..., (x_m, y_m)$ be an imitation cycle.
From this imitation cycle, we construct a strict improvement cycle as
follows: For $t = 0, ..., m-1$, we add the element $(x_t, y_{t+1})$ as
successor to $(x_t, y_t)$ and predecessor to $(x_{t+1}, y_{t+1})$. That is,
instead of simultaneous adjustments of actions at each round as in an imitation
cycle, we let players adjust actions sequentially by taking turns. The imitator adjusts from $(x_t, y_t)$ to $(x_t, y_{t+1})$ and the maximizer from $(x_{t}, y_{t+1})$ to $(x_{t+1}, y_{t+1})$ for $t = 0, ..., m-1$. This construction yields a
sequential path.

We now show that it is a strict improvement cycle. First, for the imitator,
whenever she adjusts in $t = 0, ..., m-1$, we claim $\Delta(y_t, x_t) <
\Delta(y_{t+1}, x_t) = 0$. Note that by symmetric zero-sum, $\Delta(y_t,
x_t) = - \Delta(x_t, y_t) < 0$ because $(x_t, y_t)$ is an element of an
imitation cycle, i.e., $\Delta(x_t, y_t) > 0$. $\Delta(y_{t+1}, x_t) = 0$
because the imitator mimics the action of the maximizer, $y_{t+1} = x_t$.
Thus $\Delta(y_{t+1}, x_t) = \Delta(x_t, x_t) = 0$ by symmetric zero-sum.

Second, for the maximizer, whenever she adjusts in $t = 1, ..., m$, $\Delta(x_t, y_t) > \Delta(x_{t-1}, y_t) = 0$ because $(x_t, y_t)$ is an
element of an imitation cycle, so $\Delta(x_t, y_t) > 0$. Moreover, the
imitator mimics the action of the maximizer, i.e., $y_t = x_{t - 1}$, and
thus $\Delta(x_{t-1}, y_t) = \Delta(x_{t-1}, x_{t-1}) = 0$. Hence $(x_0,
y_0), (x_0, y_1), (x_1, y_1), ..., (x_{m-1}, y_m), (x_m, y_m)$ is indeed a
strict improvement cycle. \hfill $\Box$\\

The converse is not true as the following counter-example shows.

\begin{ex}\label{ngrps_gop} Consider the following relative payoff game.\footnote{This example appears also in Ania (2008, Example 2), where it is used to demonstrate that the class of games where imitation is payoff improving (when all players are imitators) is not a subclass of generalized ordinal potential games.}
\begin{equation*}
\begin{array}{c}
\Delta =
\begin{array}{cc}
&
\begin{array}{ccc}
a & b & c%
\end{array}
\\
\begin{array}{c}
a \\
b \\
c%
\end{array}
& \left(
\begin{array}{ccc}
0 & 0 & -1 \\
0 & 0 & 1 \\
1 & -1 & 0
\end{array}
\right)
\end{array}
\end{array}
\end{equation*}
Clearly, this game is not a generalized rock-paper-scissors game. Thus, by Lemma~\ref{imitation_cycle} it does not possess an imitation cycle. However, we can construct a strict improvement cycle $(b, a)$, $(c, a)$, $(c, c)$, $(b, c)$ and $(b, a)$.
\end{ex}

\begin{prop}\label{gop} Let $(X, \pi)$ be a finite symmetric game with its relative payoff game $(X, \Delta)$. If $(X, \Delta)$ is a generalized ordinal potential game, then imitation is not subject to a money pump.
\end{prop}

\noindent \textsc{Proof. } Monderer and Shapley (1996, Lemma 2.5) show that
a finite strategic game has no strict improvement cycle (what they call the
finite improvement property) if and only if it is a generalized ordinal
potential game. Since this result holds for any finite strategic game, it
holds also for any finite symmetric zero-sum game $(X, \Delta)$.

Lemma~\ref{improvement_imitation} shows that if $(X, \Delta)$ does not
contain a strict improvement cycle, then it does not contain an imitation
cycle. Thus Lemma~\ref{imitation_cycle} implies that imitation is not
subject to a money pump. \hfill $\Box$\\

If the converse were true, then the class of generalized ordinal potential relative payoff games and relative payoff games that are not generalized rock-paper-scissors games would coincide. Yet, the converse is not true. This follows again from Example~\ref{ngrps_gop}. It is not a generalized rock-paper-scissors game but due to the existence of a strict improvement cycle it does not possess a generalized ordinal potential by Monderer and Shapley (1996, Lemma 2.5).

For an example of a game whose relative payoff game is a generalized ordinal potential game see again the coordination game with an outside option presented in Example~\ref{gopotential}. A generalized ordinal potential function is given by
\begin{equation*}
\begin{array}{c}
G =
\begin{array}{cc}
&
\begin{array}{ccc}
A & B & C
\end{array}
\\
\begin{array}{c}
A \\
B \\
C
\end{array}
& \left(
\begin{array}{ccc}
-2 & -1 & -2 \\
-1 & 0 & 0 \\
-2 & 0 & 0
\end{array}
\right)
\end{array}
\end{array}
\end{equation*}

It is straightforward to check that exact, weighted, and ordinal potential
games are generalized ordinal potential games.

\begin{cor} Let $(X, \pi)$ be a finite symmetric game with its relative
payoff game $(X, \Delta)$. If $(X, \Delta)$ is an exact potential game, a
weighted potential game, or an ordinal potential game, then imitation is not
subject to a money pump.
\end{cor}

\subsection{Quasiconcave Relative Payoff Games}

Here we show that imitation is essentially unbeatable if the relative payoff
game is ``quasiconcave''. The following definition naturally extends the
definition of quasi-concave payoff function on convex real-valued spaces to
the case of finite, symmetric games. It is the notion of single-peakedness.

\begin{defin}[Quasiconcave] A finite symmetric game $(X, \pi)$ with symmetric $m \times m$ payoff matrix $\pi =(\pi _{xy})$ is quasiconcave (or single-peaked) if for each $y \in X$, there exists a $k_{y}$ such that
\begin{equation*} \pi _{1y}\leq \pi _{2y}\leq \ldots \leq \pi _{k_{y}y} \geq \pi
_{k_{y}+1y} \geq \ldots \geq \pi _{my}.
\end{equation*}
\end{defin}

That is, we say a symmetric game is quasiconcave if each column has a single peak.
In our companion paper, Duersch, Oechssler, and Schipper (2010, Corollary 4), we show that if $X$ is finite and $\Delta$ is quasiconcave, then a fESS of $(X, \pi)$ exists.

It is clear from the definition of quasiconcavity that if $\Delta $ is
quasiconcave then there exists a total order on the action space. With some abuse of notation, we denote this order also by $\leq$. If $\Delta $ is quasiconcave, we say that $x$ \emph{is between some} $x^{\prime }$ and $x^{\prime \prime }$ if $x^{\prime} \leq x \leq x^{\prime \prime }$ or $x^{\prime \prime} \leq x \leq x^{\prime}$.

\begin{lem}\label{closerfESS} Let $(X, \Delta)$ be the relative payoff game of the symmetric game $(X, \pi)$. Suppose $\Delta$ is quasiconcave.
\begin{enumerate}
\item If $x$ is between some $y$ and some fESS $x^{\ast}$, then
\begin{equation*}
\Delta (x,y) = -\Delta(y, x) \geq 0.
\end{equation*}

\item If $x^{\ast }$ and $x^{\ast \ast }$ are fESS, then so are all $x$
between $x^{\ast }$ and $x^{\ast \ast }$.
\end{enumerate}
\end{lem}

\noindent \textsc{Proof. } (1) Let $x$ be between $y$ and $x^{\ast}$. By
definition of the fESS, $\Delta (x^{\ast },y)\geq 0.$ By symmetry of
payoffs, $\Delta (y,y)=0.$ Part (1) of the lemma follows then by
quasiconcavity.

(2) Let $x^{\ast }$ and $x^{\ast \ast }$ both be fESS. Thus, $\Delta
(x^{\ast },x^{\ast \ast })\geq 0$ and $\Delta (x^{\ast \ast },x^{\ast })\geq
0$. Since $\Delta (x^{\ast },x^{\ast \ast })=-\Delta (x^{\ast \ast },x^{\ast
})$, we must have $\Delta (x^{\ast },x^{\ast \ast })=0$. By quasiconcavity, $
\Delta (x^{\prime },x^{\ast \ast }) \geq 0$ and $\Delta (x^{\prime },x^{\ast }) \geq 0$ for all $x^{\prime}$ between $x^{\ast }$ and $x^{\ast \ast }$. By part (1)
of the lemma, $\Delta (x^{\prime }, x)\geq 0$, for all $x \in X$. Hence, all $x^{\prime }$ between $x^{\ast }$ and $x^{\ast \ast }$ are fESS as
well.\hfill $\Box$

\begin{prop}\label{qcgames} Let $(X, \pi)$ be a finite symmetric game with its relative payoff game $(X, \Delta)$. If $\Delta$ is quasiconcave, then imitation is not subject to a money pump.
\end{prop}

\noindent \textsc{Proof. } We will show that the imitator's play must reach
the set of fESS in finitely many steps, which implies that imitation is not
subject to a money pump.

We need the following notation. Let $E$ denote the (finite) set of fESS of $(X,\pi )$. By Lemma~\ref{closerfESS} (2), if $x^{\ast }$ and $x^{\ast \ast }$
are two fESS with $x^{\ast }\leq x^{\ast \ast }$, then for any $x$ with $x^{\ast }\leq x\leq x^{\ast \ast }$ also $x$ is a fESS, where the total
order $\leq $ is induced by quasiconcavity of $\Delta $. Denote $x^{\ast
}:=\min E$, $x^{\ast \ast }:=\max E$ the smallest and largest fESS,
respectively, where again the max and min are taken with respect to the
total order induced by quasiconcavity of $\Delta $. We denote by $<$ the
strict part of $\leq $, i.e., $x<x^{\prime }$ if and only if $x\leq
x^{\prime }$ and not $x^{\prime }\leq x$.

For any value $y\in X,$ $y > x^{\ast \ast }$, we define the following lower
bound (which need not always exist),
\begin{equation*} l(y):=\max \left\{ x \in X : \Delta (x,y) <0 , x < x^{\ast }\right\}.
\end{equation*}
If $y$ is larger than the largest fESS, then $l(y)$ is the largest action
lower than the lowest fESS at which relative payoffs are strictly negative.
Likewise, for $y \in X$, $y < x^{\ast}$, we define the following upper bound
(which need not always exist),
\begin{equation*}
u(y):=\min \left\{ x \in X : \Delta (x, y) < 0, x > x^{\ast \ast
} \right\}.
\end{equation*}

Without loss of generality, let $y_{0}>x^{\ast \ast }$ be the starting value
of the imitator. (The case of $y_{0}<x^{\ast }$ follows analogously.). Let
us consider all possible choices of the maximizer. By Lemma~\ref{neverworse}, the maximizer will never choose an action $x$ such that $\Delta(x, y_{0})<0$. If the maximizer chooses any $x$ such that $\Delta(x, y_{0})=0$, then the maximizer will not be imitated and the situation in $t=1$ will be identical to $t=0$. Thus, from now on we can restrict attention to $x \in X$ such that $\Delta (x, y_{0}) > 0$.

We claim that $\Delta (x,y_{0})>0$ can occur only if $x<y_{0}$ and $l(y_{0})<x$, where the second requirement is empty should $l(y_{0})$ not exist. To see that we can exclude $x \leq l(y_{0})$ note that $\Delta(l(y_{0}),y_{0})<0$ by definition. By quasiconcavity, $\Delta(x, y_0) < 0$ for all $x \leq l(y_0)$. To see that we can exclude $x \geq y_{0}$ note that $\Delta (y_{0},y_{0})=0$. By quasiconcavity, $\Delta (x,y_{0})\leq 0$, for all $x>y_{0}$. This proves the claim.

When the maximizer chooses any $x$ such that $l(y_{0})<x<y_{0}$, the
imitator imitates $x$ and chooses $y_{1}=x$ in the next period. Consider the case $y_{1}<x^{\ast }$. We claim that $u(y_{1})\leq y_{0}$. To see this note that $\Delta (y_{1},y_{0})>0$ (otherwise the imitator would not have imitated) and
hence $\Delta(y_{0}, y_{1}) < 0$. By quasiconcavity and the definition of fESS
we have $\Delta (x^{\prime },y_{1})<0$ for all $x^{\prime }\geq y_{0}$.
Hence, $u(y_{1})\leq y_{0}$. Next, consider the case that $y_{1}>x^{\ast \ast }$. In that case simply restart the procedure with the new starting value $y_{1}$.
Finally, consider the case where the maximizer selects an element in $E$.
Then no further relative payoff improvements are possible.

Thus, in this first step we have strictly narrowed down the range of the
possible choice $y_1$ of the imitator in period $t=1$ to $%
l(y_{0})<y_{1}<y_{0}$. Since $X$ is finite, when we repeat this step, the
imitator must reach the set of fESS in a finite number of steps. Once the
imitator has reached a fESS, he has reached a stationary state since then $\Delta (x,x^{\ast })\leq 0$ for all $x$. The imitator will never leave $x^{\ast }$ and the maximizer will never again obtain a positive relative payoff. Since there are only finitely many rounds in which $\Delta(x_{t},y_{t})>0$, imitation is not subject to a money pump.\hfill $\Box$\\

The following corollary may be useful for applications. Let $X \subset
\mathbb{R}^m$ be a finite subset of a finite dimensional Euclidean space. A
function $f : X \longrightarrow \mathbb{R}$ is \emph{convex (resp. concave)}
if for any $x, x^{\prime }\in X$ and for any $\lambda \in [0, 1]$ such that $\lambda x + (1 - \lambda) x^{\prime }\in X$, $f(\lambda x + (1 - \lambda)
x^{\prime }) \leq (\geq) \lambda f(x) + (1 - \lambda) f(x^{\prime })$.

\begin{cor} Let $(\mathbb{R}^m, \pi)$ be a symmetric two-player game for which $\pi(\cdot, \cdot)$ is concave in its first argument and convex in its second
argument. If the players' actions are restricted to a finite subset $X$ of
the finite dimensional Euclidian space $\mathbb{R}^m$, then imitation is not
subject to a money pump.
\end{cor}

Bargaining is an economically relevant situation involving two players. Our results imply that imitation is not subject to a money pump in bargaining as modeled in the Nash Demand game.

\begin{ex}[Nash Demand Game] Consider the following version of the Nash Demand game (see Nash, 1953). Two players simultaneously demand an amount in $\mathbb{R}_+$. If the sum is within a feasible set, i.e., $x + y \leq s$ for $s > 0$, then player 1 receives the payoff $\pi(x, y) = x$. Otherwise $\pi(x, y) = 0$ (analogously for player 2). The relative payoff function is quasiconcave. If the players' demands are restricted to a finite set, then Proposition~\ref{qcgames} implies that imitation is not subject to a money pump.
\end{ex}

\begin{ex}\label{Petergame} Consider a symmetric two-player game with the payoff
function given by $\pi (x,y)=\frac{x}{y}$ with $x,y\in X\subset \lbrack 1,2]$
with $X$ being finite. This game's relative payoff function is quasiconcave.
Thus our result implies that imitation is not subject to a money pump.
Moreover, the example demonstrates that not every quasiconcave relative
payoff function is additively separable.
\end{ex}

Finally, we like to remark that Example~\ref{ngrps_gop} is an instance of a quasiconcave relative payoff game but due to the strict improvement cycle it does not posses a generalized ordinal potential. Moreover, in Duersch, Oechssler, and Schipper (2010, Example 1) we show that there are relative payoff games that are neither generalized rock-paper-scissors games nor quasiconcave.

\subsection{Aggregative Games}

Many games relevant to economics possess a natural aggregate of all players
actions. For instance, in Cournot games the total market quantity or the
price is an aggregate. But also other games like rent-seeking games, common pool
resource games, public good games etc. can be viewed as games with an
aggregate. The aggregation property has been useful for the study of
imitation and fESS in the literature (see Schipper, 2003, and Al\'{o}s-Ferrer and Ania, 2005). In this section, we will derive results for aggregative games whose absolute payoff functions satisfy some second-order properties.\footnote{At a first glance, the aggregation property may be less compelling in the context of two-player games. However, the results we obtain in this section allows us to cover important examples that are not covered by any of our other results.}

We say that $(X,\Pi)$ is an \emph{aggregative game} if it satisfies the
following properties.

\begin{itemize}
\item[(i)] $X$ is a totally ordered set of actions and $Z$ is a totally
ordered set.

\item[(ii)] There exists an aggregator $a: X \times X \longrightarrow Z$ that is

\begin{itemize}
\item monotone increasing in its arguments, i.e. if $(x^{\prime \prime }, y^{\prime \prime }) > (x^{\prime }, y^{\prime })$, then $a(x^{\prime \prime}, y^{\prime \prime }) > a(x^{\prime }, y^{\prime })$, and

\item symmetric, i.e., $a(x, y) = a(y, x)$ for all $x, y \in X$.
\end{itemize}

\item[(iii)] $\pi$ is extendable to $\Pi :X \times Z \longrightarrow \mathbb{R}$ with $\Pi (x,a(x, y))=\pi(x, y)$ for all $x, y \in X$.
\end{itemize}

We say that an aggregative game $(X, \Pi)$ is \emph{quasisubmodular (resp.
quasisupermodular)} if $\Pi $ is \emph{quasisubmodular (resp.
quasisupermodular)} in $(x,y)$ on $X\times Z$, i.e., for all $z^{\prime
\prime }>z^{\prime }$, $x^{\prime \prime }>x^{\prime }$,
\begin{eqnarray}
\Pi (x^{\prime \prime },z^{\prime \prime })-\Pi (x^{\prime },z^{\prime
\prime })\geq 0 &\Rightarrow (\Leftarrow )&\Pi (x^{\prime \prime },z^{\prime
})-\Pi (x^{\prime },z^{\prime })\geq 0, \\
\Pi (x^{\prime \prime },z^{\prime \prime })-\Pi (x^{\prime },z^{\prime
\prime })>0 &\Rightarrow (\Leftarrow )&\Pi (x^{\prime \prime },z^{\prime
})-\Pi (x^{\prime },z^{\prime })>0.
\end{eqnarray}
Quasisupermodularity (resp. quasisubmodularity) is sometimes also called the (dual) single crossing property (e.g. Milgrom and Shannon, 1994).\footnote{It is important to realize that quasisubmodularity in $(x,z)$ where $z$ is the aggregate of \emph{all players' actions} is different from quasisubmodularity in $(x, y)$ where $y$ is the aggregate of \emph{all opponents' actions}. For instance, Schipper (2009, Lemma 1) shows that quasisubmodularity in $(x,z)$ where $z$ is the aggregate of all players' actions is satisfied in a Cournot oligopoly if the inverse demand function is decreasing. No assumptions on costs are required. It is known from Amir (1996, Theorem 2.1) that further assumptions on costs are required if the Cournot oligopoly should be quasisubmodular in $(x,y)$ where $y$ is the aggregate of all opponents'
actions.}

We say that an aggregative game $(X,\Pi )$ is \emph{submodular (resp. supermodular)} if $\Pi $ has decreasing (resp. increasing) differences in $(x,z)$ on $X\times Z$. I.e., for all $z^{\prime \prime }>z^{\prime }$, $x^{\prime \prime }>x^{\prime }$,
\begin{equation}\Pi (x^{\prime \prime },z^{\prime \prime })-\Pi (x^{\prime },z^{\prime \prime })\leq (\geq )\Pi (x^{\prime \prime },z^{\prime })-\Pi (x^{\prime },z^{\prime }).
\end{equation}
It is straight-forward to check that if an aggregative game $(X, \Pi)$ is submodular (resp. supermodular), then it is quasisubmodular (resp. quasisupermodular). The converse is false.

A finite aggregative game is \emph{quasiconcave} (or single-peaked) if for any $x, x^{\prime }, x^{\prime \prime }\in X$ with $x < x' < x''$ and $z \in Z$,
\begin{eqnarray*}
\Pi(x^{\prime }, z) \geq \min\{\Pi(x, z), \Pi(x^{\prime \prime }, z)\}.
\end{eqnarray*}
A finite aggregative game is \emph{quasiconvex} if for any $x, x^{\prime },
x^{\prime \prime }\in X$ with $x < x^{\prime }< x^{\prime \prime }$ and $z
\in Z$,
\begin{eqnarray*}
\Pi(x^{\prime }, z) \leq \max \{\Pi(x, z), \Pi(x^{\prime \prime }, z)\}.
\end{eqnarray*} It is \emph{strictly quasiconvex} if the inequality holds strictly.
An action $x^{\ast }\in X$ is a \emph{fESS} of the aggregative game
$(X,\Pi )$ if
\begin{equation*} \Pi (x^{\ast },a(x^{\ast },x))\geq \Pi (x,a(x^{\ast },x))\mbox{ for all } x \in X.
\end{equation*}

The following lemma is the key insight for our result on quasiconcave
quasisubmodular aggregative games.

\begin{lem}\label{proxESS} Suppose $(X, \Pi)$ is a quasiconcave quasisubmodular
aggregative game. If $x$ is between some $x^{\prime }$ and a fESS $x^{\ast}$, then
\begin{equation*} \Pi (x,a(x,x^{\prime }))\geq \Pi (x^{\prime }, a(x,x^{\prime })).
\end{equation*}
\end{lem}

\noindent \textsc{Proof. }Suppose that $x^{\prime }\leq x\leq x^{\ast }$.
The case $x^{\prime }\geq x\geq x^{\ast }$ can be dealt with analogously.

\noindent By the definition of a fESS

\begin{equation*}
\Pi (x^{\ast },a(x^{\ast },x^{\prime }))-\Pi (x^{\prime },a(x^{\ast
},x^{\prime }))\geq 0.
\end{equation*}

\noindent By quasiconcavity,
\begin{equation*}
\Pi (x,a(x^{\ast },x^{\prime }))-\Pi (x^{\prime },a(x^{\ast },x^{\prime
}))\geq 0.
\end{equation*}

\noindent The result follows then by quasisubmodularity,
\begin{equation*}
\Pi (x,a(x,x^{\prime }))-\Pi (x^{\prime },a(x,x^{\prime }))\geq 0,
\end{equation*}
since $a(x^*,x^{\prime })\leq a(x,x^{\prime })$. \hfill $\Box $

\begin{prop}\label{quasisub} If $(X, \Pi)$ is a finite quasiconcave quasisubmodular
aggregative game for which a fESS exists, then imitation is not subject to a
money pump.
\end{prop}

\noindent \textsc{Proof. } We will show that from any initial action, any relative payoff improving sequence of actions reaches a fESS in a finite number of steps. Once reached, there are no further improvement possibilities for the maximizer by definition of the fESS.

Note that since the game is quasiconcave, if $x^*$ and $x^{**}$ are fESS,
then so is any $x \in X$ with $x^* < x < x^{**}$ or $x^{**} < x < x^*$. We
write $E$ for the set of fESS.

\noindent \emph{Step 1}: Let $y_{0}\in X$ be the starting action of
the imitator. Assume that $y_{0}< x^{\ast } = \min E$ (the proof for $y_{0}>x^{**} = \max E$ works analogously). We claim that when the imitator switches to a new action $y_{1}\neq y_{0}$, we must have that $y_{1}>y_{0}$. Suppose by contradiction that $y_{1}<y_{0}$. By Lemma~\ref{neverworse}, the imitators would only choose $y_{1}$ if in the previous period the maximizer chose $x=y_{1}$ and received a strictly higher payoff than the imitator,
\begin{equation}\label{step1}
\Delta (y_{1},y_{0})=\Pi \left( y_{1},a(y_{1},y_{0})\right) -\Pi \left(
y_{0},a(y_{1},y_{0})\right) > 0.
\end{equation}
But this contradicts Lemma~\ref{proxESS} as $y_{1}<y_{0}<x^{\ast }$. Thus, $y_1 > y_0$.

\begin{itemize}
\item If $y_{1} \in E$, we are done.

\item If $y_{0}<y_{1}<x^{\ast }$, then take $y_{1}$ as the new starting
action and repeat Step 1.

\item Else, go to Step 2.
\end{itemize}

\noindent \emph{Step 2}: We have that $y_{1}>x^{**}$. We claim that when the
imitators switches to a new action $y_{2}\neq y_{1}$, we must have that $y_{2}<y_{1}.$ Suppose by contradiction that $y_{2}>y_{1}$. By Lemma~ \ref{neverworse}, the imitators would only choose $y_{2}$ if in the previous
period the maximizer chose $x=y_{2}$ and received a higher payoff, $\Delta
(y_{2},y_{1})>0$. But this contradicts Lemma~\ref{proxESS} as $y_{2}>y_{1}>x^{**}$. Thus $y_{2}<y_{1}.$

\begin{itemize}
\item If $y_{2} \in E$, we are done.

\item If $y_{0}<y_{2}<x^{\ast}$, then take $y_{2}$ as the new starting
action and repeat Step 1.

\item If $x^{**}<y_{2}<y_{1}$, then take $y_{2}$ as the new starting action
and repeat Step 2.
\end{itemize}

We claim that $y_{2}\leq y_{0}$ can be ruled out. Since $X$ is finite, the
algorithm then stops after finite periods. Thus the proof of the proposition
is complete once we verify this last claim.

Suppose to the contrary that $y_{2}\leq y_{0}$. By Lemma \ref{neverworse},
the imitators would only choose $y_{2}$ if in the previous period the
maximizer chose $x=y_{2}$ and received a strictly higher payoff than the imitator,
\begin{equation*}
\Delta (y_{2},y_{1})=\Pi \left( y_{2},a(y_{2},y_{1})\right) -\Pi \left(
y_{1},a(y_{2},y_{1})\right) >0.
\end{equation*}

\noindent By quasiconcavity, we have
\begin{equation*}
\Pi \left( y_{0},a(y_{2},y_{1})\right) -\Pi \left(
y_{1},a(y_{2},y_{1})\right) \geq 0.
\end{equation*}

\noindent Since $a(y_{0},y_{1}) \geq a(y_{2},y_{1})$, we have by quasisubmodularity
\begin{equation*}
\Pi \left( y_{0},a(y_{0},y_{1})\right) -\Pi \left(
y_{1},a(y_{0},y_{1})\right) \geq 0.
\end{equation*}

\noindent But this contradicts inequality~(\ref{step1}) and proves the claim. \hfill $\Box$\\

The following examples present applications of the previous result. The first example extends the linear Cournot oligopoly of Example~\ref{linear_Cournot} to general symmetric Cournot oligopoly.

\begin{ex}[Cournot Duopoly]\label{Cournot} Let the symmetric payoff function be $\pi (x,y)=xp(x+y)-c(x)$ and assume that $\pi (x,y)$ is quasiconcave in $x$. Schipper (2009, Lemma 1) shows that a symmetric Cournot duopoly with an arbitrary decreasing inverse demand function $p$ and arbitrary increasing cost function $c$ is an
aggregative quasisubmodular game. Thus, Proposition~\ref{quasisub} implies
that imitation is not subject to a money pump in Cournot duopoly.
\end{ex}

\begin{ex}[Rent Seeking]\label{rentseeking} Two contestants compete for a rent $v>0$ by bidding $x,y\in X\subseteq \mathbb{R}_{+}$. A player's probability of winning is
proportional to her bid, $\frac{x}{x+y}$ and zero if both players bid zero.
The cost of bidding equals the bid. The symmetric payoff function is given
by $\pi (x,y)=\frac{x}{x+y}v-x$ (see Tullock, 1980, and Hehenkamp,
Leininger, and Possajennikov, 2004). This game is an aggregative
quasisubmodular game (see Schipper, 2003, Example 6, and Al\'{o}s-Ferrer and
Ania, 2005, Example 2) and $\pi (x,y)$ is concave in $x$. Thus Proposition~\ref{quasisub} implies that imitation is not subject to a money pump.
\end{ex}

For quasiconvex quasisupermodular aggregative games we can prove an
analogous result. We first observe that in a strictly quasiconvex quasisubmodular game a fESS must be a ``corner'' solution if it exists. It follows that there can be at most two fESS.

\begin{lem}\label{corner_fESS} Let $(X, \Pi)$ be a finite strictly quasiconvex
quasisupermodular aggregative game. If $x^*$ is a fESS, then $x^* = \max X$ or $x^* = \min X$.
\end{lem}

\noindent \textsc{Proof. } Let $x^*$ be a fESS and suppose to the contrary that
there exist $x^{\prime }, x^{\prime \prime }\in X$ such that $x'< x^* <
x''$. We distinguish four cases:

\emph{Case 1:} If
\begin{equation*}
\Pi (x^{\prime \prime },a(x^{\ast },x^{\prime \prime }))\geq \Pi (x^{\prime
},a(x^{\ast },x^{\prime \prime })),
\end{equation*}%
then by strict quasiconvexity
\begin{equation*}
\Pi (x^{\ast },a(x^{\ast },x^{\prime \prime }))<\Pi (x^{\prime \prime
},a(x^{\ast },x^{\prime \prime })),
\end{equation*}
a contradiction to $x^{\ast }$ being a fESS.

\emph{Case 2:} The case $\Pi (x^{\prime },a(x^{\ast },x^{\prime }))\geq \Pi
(x^{\prime \prime },a(x^{\ast },x^{\prime }))$ is analogous to Case 1.

\emph{Case 3:} If
\begin{equation*}
\Pi (x^{\prime },a(x^{\ast },x^{\prime \prime }))\geq \Pi (x^{\prime \prime
},a(x^{\ast },x^{\prime \prime })),
\end{equation*}%
then by strict quasiconvexity
\begin{equation*}
\Pi (x^{\ast },a(x^{\ast },x^{\prime \prime }))<\Pi (x^{\prime },a(x^{\ast
},x^{\prime \prime })).
\end{equation*}%
By quasisupermodularity,
\begin{equation*}
\Pi (x^{\ast },a(x^{\ast },x^{\prime }))<\Pi (x^{\prime },a(x^{\ast
},x^{\prime })).
\end{equation*}%
a contradiction to $x^{\ast }$ being a fESS.

\emph{Case 4:} The case $\Pi (x^{\prime \prime },a(x^{\ast },x^{\prime
}))\geq \Pi (x', a(x^{\ast },x^{\prime }))$ is analogous to Case 3.

Thus, if $x^*$ is a fESS, then $x^* = \max X$ or $x^* = \min X$.\hfill $\Box$

\begin{prop}\label{quasisuper} If $(X, \Pi)$ is a finite strictly quasiconvex
quasisupermodular aggregative game for which a fESS exists, then imitation
is not subject to a money pump.
\end{prop}

\noindent \textsc{Proof. } We show that a process of steps that strictly increase the sum of the maximizer's relative payoffs must lead to a fESS. Note that only nontrivial steps, in which the maximizer does not repeat her action, can improve the sum of her relative payoffs. Consider a sequence of nontrivial steps $x_{1}, x_{2}, x_{3}$ the maximizer may take. Suppose that $x_{2} < x_{1}$ (the case $x_{2}>x_{1}$ is dealt with analogously). By Lemma~\ref{neverworse} it must hold that
\begin{equation}\label{star}
\Pi(x_{2}, a(x_{2},x_{1})) \geq \Pi(x_{1}, a(x_{2},x_{1})).
\end{equation}
To show that the process moves to one of the corners, we need to show
that either $x_{3} > x_{1}$ or $x_{3} < x_{2}$. Suppose to the contrary that $x_{2} < x_{3} \leq x_{1}$.\footnote{The case of $x_2 \neq x_3$ is already excluded by the requirement of non-trivial steps.} By Lemma~\ref{neverworse} it must hold that
$$\Pi (x_{3},a(x_{3},x_{2}))\geq \Pi (x_{2},a(x_{3},x_{2})).$$

Thus, by quasisupermodularity
\begin{equation}\label{fromquasi}
\Pi (x_{3},a(x_{1},x_{2}))\geq \Pi (x_{2},a(x_{1},x_{2})).
\end{equation}

From inequality~(\ref{star}) follows by strict quasiconvexity that
\begin{equation} \label{fromstar}
\Pi (x_{2},a(x_{2},x_{1}))\geq \Pi (x_{3},a(x_{2},x_{1})),
\end{equation}
with equality only for $x_{3}=x_{1}$. Thus for $x_{3}<x_{1}$, inquality~(\ref{fromstar}) yields a contradiction to inequality~(\ref{fromquasi}). For $x_{3}=x_{1}$ the sequence of steps has not improved the maximizer's sum of relative payoffs since both players obtained the same payoff throughout.

Thus we have shown that with every nontrivial step, the maximizer gets closer to a corner. Since there are only finitely many actions, if the sequence of actions is non-constant, then a corner must be reached in finitely many steps. If the corner is a fESS, then no further changes of actions occur. Otherwise, the other corner may be reached in one additional step. This must be a fESS by Lemma~\ref{corner_fESS} since a fESS is assumed to exist. Once it is reached, no further changes of actions occur.\hfill $\Box$

\section{Summary and Discussion\label{discussion}}

In Table~\ref{summary} we summarize our results.\footnote{More results on the classes of games and their relationships are contained in our companion paper, Duersch, Oechssler, and Schipper (2010).} The only class of symmetric games in which imitation can really be beaten is the class of games whose relative payoff function is a generalized rock--paper--scissors game. While this is a generic class of symmetric games, so is its complement. More importantly, many economically relevant games are contained in this complement. Thus it is fair to say that imitation seems very hard to beat in large classes of economically relevant and generic games.
\begin{table}[h!]
\caption{Summary of Results}\label{summary}
\begin{center}
\footnotesize
\begin{tabular}{llll} \hline\hline
&  &  &  \\
\textbf{Class} & \textbf{Result} & \textbf{Reference} & \textbf{Examples} \\
&  &  &  \\ \hline &  &  &  \\
Symmetric 2x2 games & essentially unbeatable & Prop.~\ref{2x2} & Chicken, Prisoners' Dilemma, \\
& &  & Stag Hunt \\ &  &  &  \\ \hline &  &  &  \\
Additively separable relative & essentially unbeatable & Prop.~\ref{separable} &  Linear Cournot duopoly \\
payoff function & &  & Heterogeneous Bertrand duopoly \\
or &  &  & Public goods \\
Relative payoff functions & essentially unbeatable & Cor.~\ref{order} &  Common pool resources \\
with increasing or decreasing & &  &  Minimum effort coordination \\
differences &  &  &  Synergistic relationship \\
or &  &  & Arms race \\
Relative payoff games with & essentially unbeatable & Cor. \ref{exact} &  Diamond's search \\
exact potential & &  &\\  &  &  &  \\ \hline &  &  &  \\
Relative payoff games with & no money pump & Prop.~\ref{gop} & Example \ref{gopotential} \\
generalized ordinal potential & &  &  \\ &  &  &  \\
Quasiconcave relative & no money pump & Prop.~\ref{qcgames} & Nash demand game \\
payoff games & & & Example~\ref{ngrps_gop} \\ &  &  & Example~\ref{Petergame} \\ &  &  &  \\
Quasiconcave quasisub- & no money pump & Prop. \ref{quasisub} & Cournot games \\
modular aggregative games & &  & Rent seeking \\ &  &  &  \\
Quasiconvex quasisuper- & no money pump & Prop.~\ref{quasisuper} &  \\
modular aggregative games & &  &  \\ &  &  &  \\ \hline &  &  &  \\
No generalized  & no money pump & Thm.~\ref{characterization} & all of the above \\
Rock-Paper-Scissors games & &  &  \\ &  &  &  \\ \hline
\end{tabular}
\end{center}
\normalsize
\end{table}

However, one needs to be aware of the limitations of our analysis, primarily the restriction to two--player games. While a full treatment of the $n$--player case is beyond the scope of the current paper, we provide here an example that shows how imitation can be beaten in a standard Cournot game when there are three players. Let the inverse demand function be $p(Q)=100 - Q$ and the cost function be $c(q_{i}) = 10 q_{i}$. Now consider the case of two maximizers and one imitator. Writing a vector of quantities as $(q_{I}, q_{M}, q_{M})$, it is easy to check
that the following sequence of action profiles $(0, 22.5, 22.5), (22.5, 0, 68), (0, 22.5, 22.5), (22.5,68,0), (0,22.5,22.5) ...$ is an imitation cycle. The two maximizers take turns in inducing the imitator to reduce his quantity to zero by increasing quantity so much that price is below marginal cost. Since the other maximizer has zero losses, she is imitated in the next period, which yields half of the monopoly profit for both maximizers. Clearly, this requires coordination among the two maximizers but this can be achieved in an infinitely repeated game by the use of a trigger strategy. Thus, imitation is subject to a money pump. Recall, however, that we pitted imitation against truly sophisticated
opponents. Whether imitation can be beaten also by less sophisticated (e.g.
human) opponents remains to be seen in future experiments.

Furthermore, our analysis was based on the assumption that an imitator sticks to his action in case of a tie in payoffs. To see what goes wrong with an alternative tie-braking rule consider a homogenous Bertrand duopoly with constant marginal costs. Suppose the imitator starts with a price equal to marginal cost. If the maximizer chooses a price strictly above marginal cost, her profit is also zero. If nevertheless, the maximizer were imitated, she could start the money pump by undercutting the imitator until they reach again price equal to marginal cost and then start the cycle again. This example shows that our results depend crucially on the details of the imitation heuristics. It would be interesting to exactly characterize the class of simple decision heuristics that are essentially unbeatable in large classes of economically relevant games. We leave this for future work.

\end{document}